\documentclass[preprint,12pt,authoryear]{elsarticle}

\usepackage{amssymb}
\usepackage{amsmath}

\journal{Applied Mathematics and Computation}

\begin{document}

\begin{frontmatter}



\title{Periodic boundary conditions on staggered grids: Uniqueness of variables at cell edges/faces} 


\author{Divyanshu Gola$^{a}$} 

\affiliation{organization={Department of Mechanical and Aerospace Engineering, University of California San Diego},
            addressline={9500 Gilman Dr}, 
            city={La Jolla},
            postcode={92093}, 
            state={CA},
            country={USA}}

\begin{abstract}
Periodic boundary conditions when applied to staggered grids, which define variables on both cell edges and cell centers, can be shown to have a problem with uniqueness of variables at cell edges depending on the number of points in the direction of periodicity. In the context of the grid defined in this work, it can be shown that uniqueness is guaranteed if and only if the number of points in the periodic direction are odd. This stems from the rank of the matrix with dimensions $(N-2) \times (N-2)$ that transforms the values at cell centers to values at edges. This matrix is full rank if and only if $N$ is odd. Here, $N$ is the number of points describing the cell edges.
\end{abstract}






\end{frontmatter}



\section{Introduction}
\label{sec1}

\begin{figure}[t]
\centering
\centerline{\includegraphics[width=0.8\linewidth, keepaspectratio]{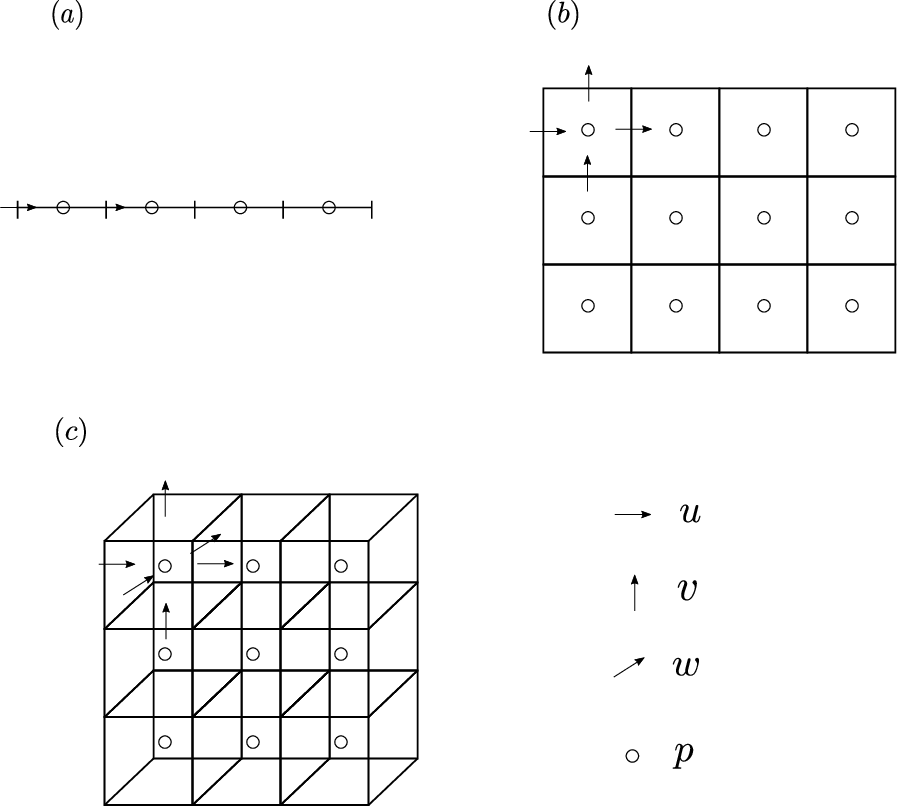}}
\caption{Examples of staggered grids in one, two,  and three dimensions. Here, pressure ($p$) is the centered while the  components of velocity are staggered on edges/faces.}\label{fig1}
\end{figure}

Periodic boundary conditions are used in a multitude of numerical solvers aimed at computationally solving a system of differential equations by discretising them in space and time using finite differences. Some applications include simulating compressible turbulence in a shear flow \citep{sarkar_direct_1991}, wave propagation \citep{rivera_bed_2017} and even equations in cylindrical coordinates that allow for the azimuthal direction to be periodic due to axisymmetry \citep{ortiz_stratified_2019}. The periodicity along one or more directions can be leveraged numerically via Fourier and cyclic reduction methods \citep{hockney_fast_1965, swarztrauber_methods_1977, rossi_parallel_1991} in that direction to allow for faster computation times. 

Staggered grids are used in simulations to avoid decoupling between variables such as pressure and velocity. This is done by offsetting velocities from the centers to the edges of the cell so that second oder accurate gradients can be defined on the cell centers, (which also stores pressure and density) without decoupling information on alternating cells which can lead to checkerboard instability. Figure \ref{fig1} shows examples of staggered grids in one, two and three dimensions.

When using periodic boundary conditions with staggered grids, the value of a staggered variable at the center is approximated to be the mean of the values at the edges. This is locally second order accurate just like the central finite difference of the first derivative that is calculated on the center using the edges. Calculating the values of a centered variable at the edges is not as straightforward as just taking the mean of the values at the centers and one needs to look at the system of equations that relate them. 

The problem of determining the values at edges given the value of centers on a staggered grid is the highlight of this paper. It is formally described in one dimension in section \ref{desc} and a couple of examples are shown. Section \ref{gen} generalizes to any number of points in the periodic direction. Implications and conclusions are presented in section \ref{conc}.

(Note that for a 3D cell, `cell face' is the appropriate term but the author will primarily use the term `cell edge' to refer to both edges and faces unless stated otherwise)


\section{Problem description and examples}
\label{desc}


\begin{figure}[t]
\centering
\centerline{\includegraphics[width=0.9\linewidth, keepaspectratio]{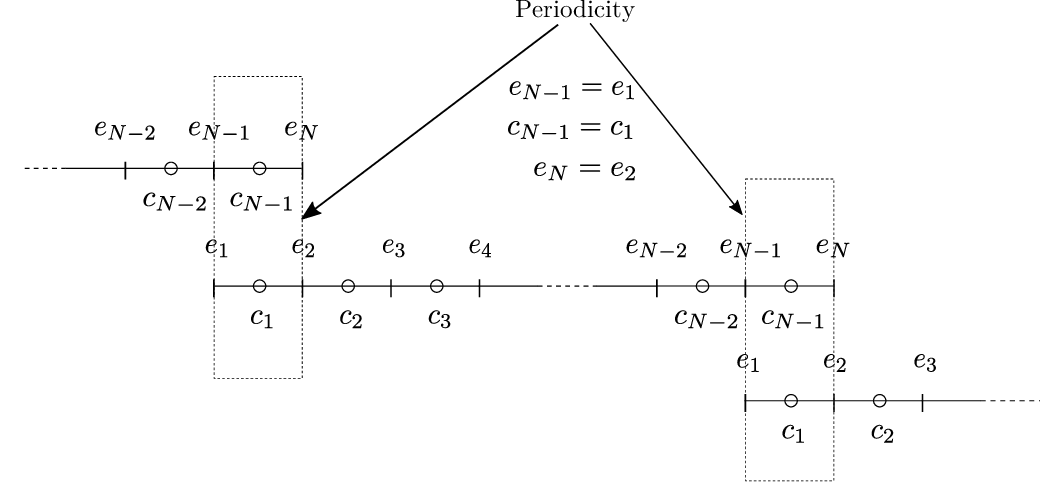}}
\caption{Problem description on one dimensional staggered grid with periodicity: Given all $c_{i}$, the goal is to find all $e_{i}$ so that $c_{i} = (e_{i}+e_{i+1})/2$.}\label{fig2}
\end{figure}

The problem is posed in one dimenion for simplicity as follows: Consider $N$ arbitrarily spaced points $i = 1,2,3,...,N$ with $N > 2$  and let these denote the cell edges. For any two consecutive points, we describe the cell centers as $i+1/2, i = 1,2,3,...(N-1)$. Also, let $\phi$ be a variable which takes values $e_{i}$ and $c_{i}$ on edge $i$ and center $i+1/2$ respectively so that $c_{i} = (e_{i}+e_{i+1})/2$ Enforcing periodicity , we get $e_{N-1} = e_{1}, e_{N} = e_{2}$ and $c_{N-1} = c_{1}$ (Figure \ref{fig2}). The goal is to find all $e_{i}$ if all $c_{i}$ are known. Two example cases, $N = 5$ and $N = 6$ are considered and then a generalisation for any $N$ is presented.

\subsubsection{Example case $N = 5$}

For $N=5$, we get the following system of equations:

\begin{align}
e_1 + e_2= 2c_1 \\
 e_2 + e_3 =2c_2\\
 e_3+ e_4=2c_3\\
 e_4 + e_5=2c_4
 \end{align}

Using periodicity, (4) and (1) are the same equations. Also substituting $e_4=e_1$ in (3) gives:

\begin{align}
e_1 + e_2= 2c_1 \\
 e_2 + e_3 =2c_2\\
 e_3+ e_1=2c_3
 \end{align}

In matrix form (5),(6) and (7) can be expressed as 

\begin{equation}
\begin{bmatrix}
    1 & 1 & 0 \\
    0 & 1 & 1 \\
    1 & 0 & 1
\end{bmatrix}
\begin{bmatrix}
    e_1\\
    e_2\\
    e_3
\end{bmatrix}
=
\begin{bmatrix}
    2c_1\\
    2c_2\\
    2c_3
\end{bmatrix}
\end{equation}

The matrix on the left hand side of (8) has a determinant of  $2 (\neq 0)$ and therefore is invertible. This implies that the system has a unique solution or that the values at the edges can be uniquely determined from the values at the centers. The solution is $e_1 = c_1 - c_2 +c_3, e_2 = c_1 + c_2 - c_3, e_3 = -c_1 + c_2 + c_3$.

\subsubsection{Example case $N = 6$}

For $N=6$, we get the following system of equations (after using periodicity):

\begin{align}
e_1 + e_2= 2c_1 \\
 e_2 + e_3 =2c_2\\
 e_3+ e_4=2c_3\\
 e_4 + e_1=2c_4
 \end{align}

In matrix form (9),(10),(11) and (12) can be expressed as 

\begin{equation}
\begin{bmatrix}
    1 & 1 & 0 & 0\\
    0 & 1 & 1 & 0\\
    0 & 0 & 1 & 1\\
    1 & 0 & 0 & 1
\end{bmatrix}
\begin{bmatrix}
    e_1\\
    e_2\\
    e_3\\
    e_4
\end{bmatrix}
=
\begin{bmatrix}
    2c_1\\
    2c_2\\
    2c_3\\
    2c_4
\end{bmatrix}
\end{equation}

The matrix on the left hand side of (13) is singular which can be seen from the row echelon form,

\begin{equation}
\begin{bmatrix}
    1 & 1 & 0 & 0\\
    0 & 1 & 1 & 0\\
    0 & 0 & 1 & 1\\
    0 & 0 & 0 & 0
\end{bmatrix}
\begin{bmatrix}
    e_1\\
    e_2\\
    e_3\\
    e_4
\end{bmatrix}
=
\begin{bmatrix}
    2c_1\\
    2c_2\\
    2c_3\\
    2c_4 - 2c_3 + 2c_2 - 2c_1
\end{bmatrix}
\end{equation}

\begin{figure}[t]
\centering
\centerline{\includegraphics[width=0.9\linewidth, keepaspectratio]{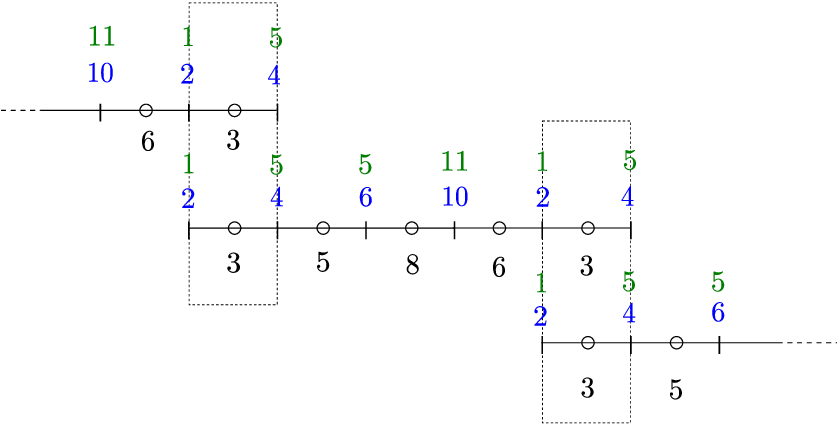}}
\caption{Example case $N=6$ showing more than one possibility for values at edges (colored green and blue) when values at centers (colored black) are given.}\label{fig3}
\end{figure}

Therefore, the system of equations (13) has infnitely many solutions if $c_4 = c_3 - c_2 + c_1$ and no solution otherwise. 

The final row of zeros is obtained because of the even number of edges in the problem ($N=6$) and therefore an even number of dimension of the matrix ($N-2 = 4$). Figure \ref{fig3} shows an example for $N = 6$ where two different set of values at the edges give the same values at the centers.

\section{Generalisation}
\label{gen}
Consider the problem for $N$ edges and $N-1$ centers (Figure \ref{fig2}). Periodicity implies $e_{N-1} = e_{1}, e_{N} = e_{2}$ and $c_{N-1} = c_{1}$. Then, the system of equations in matrix form is 

\begin{equation}
\begin{bmatrix}
    1 & 1 & 0 & \dots  & 0 & 0\\
    0 & 1 & 1 & \dots  & 0 & 0\\
    0 & 0 & 1 & \dots  & 0 & 0\\
    \vdots & \vdots & \vdots & \ddots & \vdots & \vdots \\
    0 & 0 & 0 & \dots  & 1 & 1 \\
    1 & 0 & 0 & \dots  & 0 & 1
\end{bmatrix}
\begin{bmatrix}
    e_1\\
    e_2\\
    e_3\\
    \vdots\\
    e_{N-3}\\
    e_{N-2}
\end{bmatrix}
=
\begin{bmatrix}
    2c_1\\
    2c_2\\
    2c_3\\
    \vdots\\
    2c_{N-3}\\
    2c_{N-2}
\end{bmatrix}
\end{equation}

which in row echelon form becomes

\begin{equation}
\begin{bmatrix}
    1 & 1 & 0 & \dots  & 0 & 0\\
    0 & 1 & 1 & \dots  & 0 & 0\\
    0 & 0 & 1 & \dots  & 0 & 0\\
    \vdots & \vdots & \vdots & \ddots & \vdots & \vdots \\
    0 & 0 & 0 & \dots  & 1 & 1 \\
    0 & 0 & 0 & \dots  & 0 & 1-(-1)^{N-2}
\end{bmatrix}
\begin{bmatrix}
    e_1\\
    e_2\\
    e_3\\
    \vdots\\
    e_{N-3}\\
    e_{N-2}
\end{bmatrix}
=
\begin{bmatrix}
    2c_1\\
    2c_2\\
    2c_3\\
    \vdots\\
    2c_{N-3}\\
  2 \sum_{i=1}^{N-2} (-1)^{N-2-i}c_{i}
\end{bmatrix}
\end{equation}

The matrix on the left hand side in (16) has determinant 2 when $N$ is odd (full rank matrix with rank = $N-2$) and determinant 0 when $N$ is even (rank = $N-3$). Evidently, the values at cell centers can be uniquely determined in the former case while the latter case will have infinitely many solutions if $c_{N-2} = \sum_{i=1}^{N-2-1} (-1)^{N-2-i}c_{i}$ and no solution otherwise.

\section{Implications and conclusion}
\label{conc}

The analysis in section \ref{gen} shows that a variable under periodic boundary conditions on a staggered grid can admit unique, infinitely many or no solutions for its values on the edges when the values at the centers are known. For example, going back to incompressible and compressible flow solvers, if pressure is known at centers from a simulation and staggered values of pressure are needed at the edges/faces in the direction that is periodic (say to compute forces on that edge/face) the values can only be determined uniquely if the number of edges described by the grid are odd. For an even number of points describing the edges, this problem can be avoided by calculating the values at the edges by using other interpolation techniques or by solving the system of equations on an odd number of points by assuming an appropriate value on the first or the third last point.

The author suggests that this result be looked at as more of a mathematical result rather than a computational one because as far as the author knows, having even number of grid points describing the edges does not seem to affect the numerical stability of the solver itself.




\bibliographystyle{elsarticle-harv} 
\bibliography{periodic_stagger_v1}






\end{document}